\documentclass{article}

\usepackage{arxiv}

\usepackage[utf8]{inputenc} 
\usepackage[T1]{fontenc}    
\usepackage[hidelinks]{hyperref}       
\usepackage{url}            
\usepackage{booktabs}       
\usepackage{amsfonts,amsmath}       
\usepackage{nicefrac}       
\usepackage{microtype}      
\usepackage{lipsum}
\usepackage{graphicx}
\usepackage{bm}
\graphicspath{ {./images/} }

\newcommand\blfootnote[1]{%
  \begingroup
  \renewcommand\thefootnote{}\footnote{#1}%
  \addtocounter{footnote}{-1}%
  \endgroup
}

\title{Assessing the dynamic response of long-span bridges under simultaneous wind and traffic loads}

\author{
 Gledson Rodrigo Tondo$^{\star}$ \\
  Bauhaus-Universität Weimar\\
  Weimar, Germany \\
  \And
 Guido Morgenthal \\
  Bauhaus-Universität Weimar\\
  Weimar, Germany \\
}

\begin{document}
\maketitle
\begin{abstract}
Wind-traffic interactions strongly influence the dynamic response of long-span bridges, yet loads are often analysed independently. This work models concurrent wind and traffic and demonstrates that it differs from linear superposition. Traffic is synthesised from volumes, composition, and vehicle dynamics, with vehicles represented as 3D systems. Vehicle-pavement interaction adopts ISO roughness with transverse coherence, and wind turbulence follows the Kármán spectrum with Davenport coherence. A quasi-steady aerodynamic model supports time-history analysis under combined actions. Results indicate non-linear interactions that change response, revealing limitations of conventional design assumptions. The framework enables accurate performance assessment and informs serviceability criteria and design optimisation for long-span bridges.
\end{abstract}


\section{Introduction}\blfootnote{IABSE Symposium Copenhagen 2026 - Bridging Advanced Technologies - Structural Innovation\\ \hspace*{5mm} Copenhagen, DK, 2026}

Long-span bridges are uniquely sensitive to dynamic environmental and live loads. Wind, in particular, can induce significant vibrations in flexible spans, while heavy traffic can cause oscillations and deflections through moving vehicular loads~\cite{chen2010,camara2019,wang2014}. Historically, design analyses consider wind and traffic effects separately, with wind-induced responses evaluated without vehicles on the bridge, and traffic load effects are assessed under assumed calm air conditions~\cite{chen2010,xu2003}. In reality, however, these loads may act simultaneously: vehicles cross bridges during strong winds, and their interaction can be complex. Wind gusts can affect vehicle distribution of loads and bridge aerodynamics, while vehicle mass and damping can alter the bridge’s dynamic response and therefore also the wind response. Recent studies have highlighted the importance of treating the bridge, vehicles, and wind as a coupled system to predict dynamic behaviour more accurately~\cite{zhang2024,xu2003}. There is a need for integrated analysis to capture such non-linear load coupling effects. Alternatively, these loads can be investigated in practice through load reconstruction techniques~\cite{tondo2024,tondo2025}.

This paper presents a framework for simulating the combined wind and traffic loading on long-span bridges and assessing the resulting dynamic response. A case study is carried out on the Great Belt East Bridge, to quantify differences between separate and simultaneous load analysis. In the following, Section~\ref{sec:2} describes the modelling approach for structural dynamics, aerodynamic loading, and traffic loading. Section~\ref{sec:3} presents the dynamic analysis, defining simulation scenarios and discussing results from the combined wind-traffic simulations. Finally, Section~\ref{sec:4} draws conclusions regarding the implications of simultaneous loading for bridge design and performance assessment.

\section{Modelling approach} \label{sec:2}

\subsection{Structural modelling}

\begin{figure}[h]
\centering
\includegraphics[width=10cm]{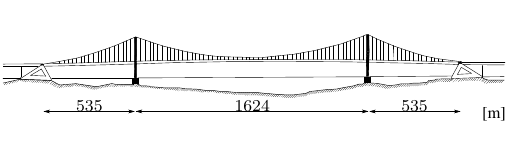}\\
\includegraphics[width=10cm]{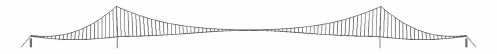}\\
\includegraphics[width=10cm]{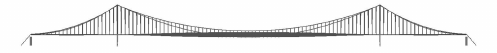}\\
\caption{The Great Belt East Bridge: schematic (top), with the first bending mode (centre, $f_h = 0.100$ Hz) and first torsional mode (bottom, $f_\alpha = 0.278$ Hz).}
\label{fig:01}
\end{figure}

\begin{figure}[h]
\centering
\includegraphics[width=10cm]{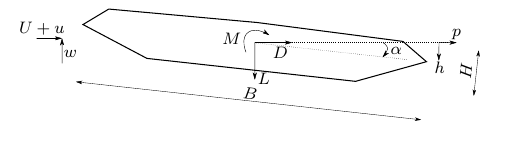}
\caption{Cross-section ($B$, $H$) and degrees of freedom ($h$, $p$, $\alpha$), mean and turbulent wind components ($U$, $u$, $w$), and aerodynamic forces ($D$, $L$, $M$).}
\label{fig:02}
\end{figure}

The Great Belt Bridge is a suspension bridge with a main span of 1624 m and side spans of 535 m (see Figure~\ref{fig:01} for a schematic of the bridge and the first two mode shapes)~\cite{tondo2024}. The steel box-girder deck is supported by two main cables and vertical hangers. Figure~\ref{fig:02} illustrates the deck cross-section and coordinate system ($h$, $p$ and $\alpha$) for aerodynamic forces and structural DOFs. The sectional properties - deck width $B$=31 m and depth $H$=4.30 m, along with the along-wind mean and turbulence velocities $U$ and $u$ and vertical turbulence $w$, and aerodynamic forces (drag $D$, lift $L$ and moment $M$) are also indicated. 

The modal properties of the structure are obtained from a calibrated finite element model. A total of 30 modes is considered in the analysis. A modal damping model is assumed, with a damping ratio of 0.5\% to represent structural damping in calm air. The equations of motion of the bridge in modal coordinates $q(t)$ can be written as

\begin{equation}
M\ddot{q} + C\dot{q} + Kq = F_{\mathrm{wind}}(t) + F_{\mathrm{veh}}(t),
\end{equation}

where $M$, $C$ and $K$ are the modal mass, damping, and stiffness matrices, and $F_{\mathrm{wind}} (t)$ and $F_{\mathrm{veh}} (t)$ are the generalised force vectors due to wind and vehicles, respectively. These forces are described in the following sections.

\subsection{Aerodynamics}

Wind loading on the bridge is modelled using a turbulent buffeting approach. A mean wind speed $U$ is assumed perpendicular to the bridge span. Superimposed on the mean wind are turbulent fluctuations in all three directions: along-wind $u(t)$, vertical $w(t)$, and lateral $v(t)$. These wind velocity fluctuations are simulated as Gaussian random processes with prescribed spectral properties. The turbulence is characterised by a von Kármán power spectral density (PSD), which defines how energy is distributed across frequencies~\cite{tondo2024}. The PSD is defined by turbulence lengthscale and intensity parameters in accordance with wind climate data from the site~\cite{kavrakov2024}. 

\begin{figure}[h]
\centering
\includegraphics[]{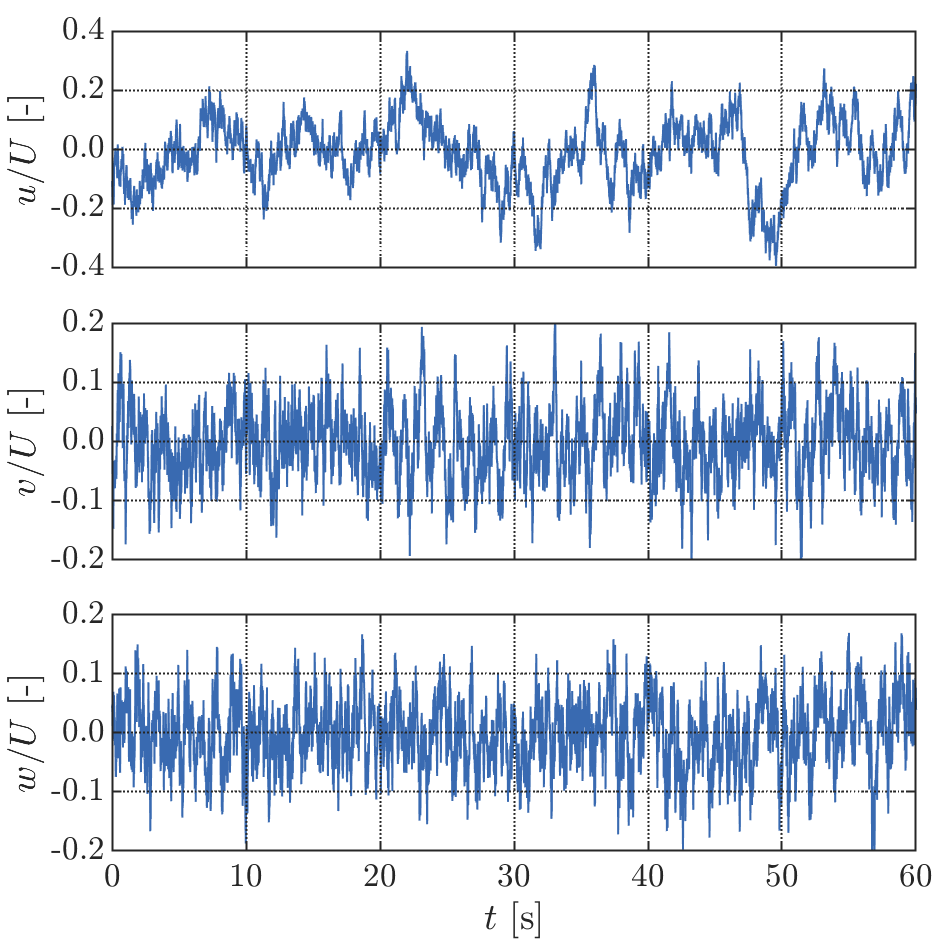}
\caption{Excerpts of the turbulent wind signal at midspan, in the along-wind direction (top), along-span direction (centre) and vertical direction (bottom).}
\label{fig:03}
\end{figure}

The turbulent wind time series are generated at multiple positions along the span. Davenport’s exponential coherence model is used to impose realistic spatial correlation of the turbulent field along the span. A Fourier-based simulation is employed to produce samples of $u(t)$, $v(t)$, $w(t)$ consistent with the target spectra and coherences. Figure~\ref{fig:03} shows an example of a $60~\mathrm{s}$ excerpt of the simulated wind velocity fluctuations at midspan in the along-wind, lateral, and vertical directions, normalised by the mean wind speed $U$.

Aerodynamic forces on the deck are computed using a quasi-steady approach~\cite{tondo2023}, as

\begin{align}
D &= F_L \sin \phi_D - F_D \cos \phi_D, \\
L &= F_L \cos \phi_L - F_D \sin \phi_L, \\
M &= F_M,
\end{align}

with $\phi_i$, $i \in \{D,L,M\}$ is the dynamic angle of attack, and

\begin{align}
F_D &= \frac{1}{2}\rho U_r^2 B C_D(\alpha_e), \\
F_L &= \frac{1}{2}\rho U_r^2 B C_L(\alpha_e), \\
F_M &= \frac{1}{2}\rho U_r^2 B^2 C_M(\alpha_e),
\end{align}

where $\rho$ is the air density, $C_i$ is the static wind coefficient and $\alpha_{e i}$ the effective angle of attack, calculated as

\begin{equation}
\alpha_e = \alpha_s + \alpha + \arctan\left(\frac{w+\dot{h}+m_i B \dot{\alpha}}{U+u-\dot{p}}\right),
\end{equation}

where $\alpha_s$ is the static angle of attack and $m_i$ controls the aerodynamic centre. The resultant velocity $U_{r i}$ is calculated as

\begin{equation}
U_{ri} = \sqrt{(w+\dot{h}+m_i B \dot{\alpha})^2 + (U+u-\dot{p})^2}.
\end{equation}

\subsection{Traffic}

\subsubsection{Dynamic vehicle models}

\begin{figure}[h]
\centering
\includegraphics[width=5.5cm,trim=9.6cm 0cm 0cm 0cm,clip]{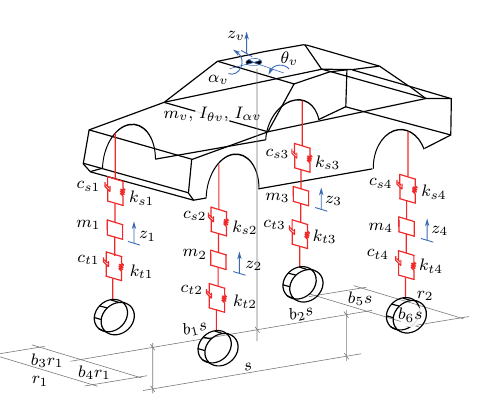}
\includegraphics[width=10.5cm]{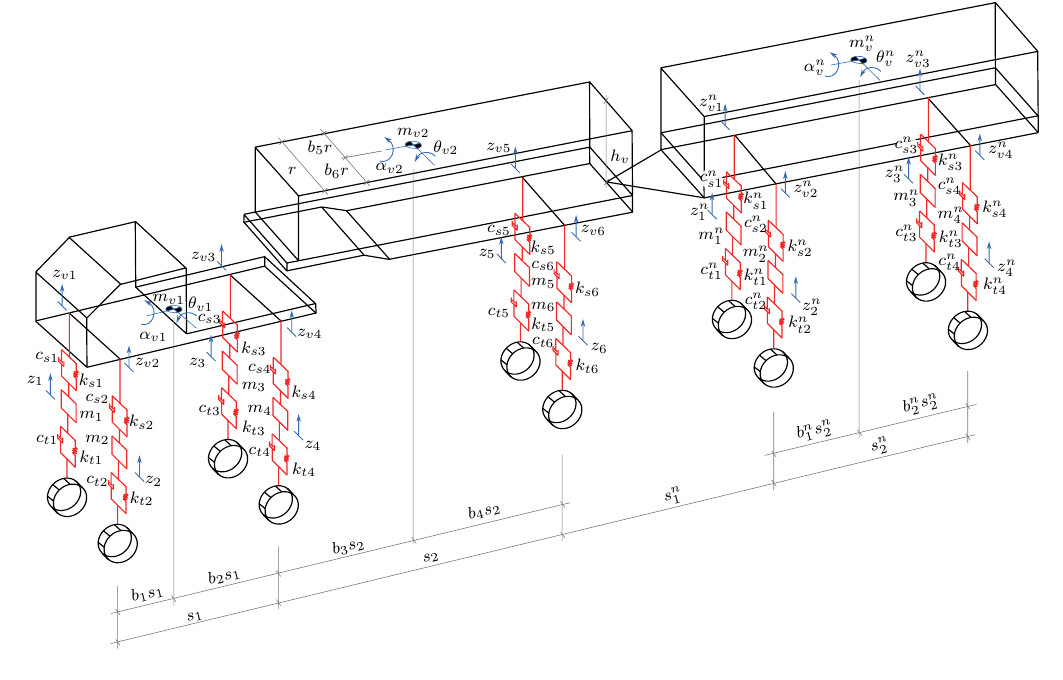}
\caption{Examples of 2-axle and 3-axle dynamic vehicle models used in the traffic analysis.}
\label{fig:04}
\end{figure}

Each vehicle travelling on the bridge is represented as a multi-degree-of-freedom dynamic system that captures its vertical and rotational motions. Several classes of vehicles are modelled, including passenger cars, vans, buses, and articulated trucks. Figure~\ref{fig:04} depicts simplified models for two vehicle types along with their coordinate system and DOFs. The left diagram illustrates a two-axle model with a sprung mass $m_v$ and pitch and roll inertias $I_{\theta v}$ and $I_{\alpha v}$, having a vertical displacement DOF $z_v$, as well as pitch and roll ($\theta_v$ and $\alpha_v$) motions, since the model accounts for both left and right wheel paths. The car’s suspension is represented by spring-damper pairs ($k_{s n}$, $c_{s n}$ for $n \in \{1,\dots,4\}$), and each wheel assembly $m_n$ is connected via a tyre stiffness $k_{t n}$ and damping $c_{t n}$. The right diagram in Figure~\ref{fig:04} shows the model for an articulated truck. They similarly include a sprung mass with vertical, pitch and roll DOFs, and unsprung masses for each axle or wheel set. The remaining parameters $r_n$, $s_n$ and $b_n$ are related to the geometrical properties of each vehicle.

The values for each of the vehicle parameters are obtained from the literature~\cite{fafard1997,tian2020,xu2003}. To account for variability in the traffic properties, each vehicle moving in the bridge has its unsprung mass sampled from a uniform distribution, with a standard deviation equivalent to 5\% of the mean value.

The vehicles are assumed to have linear suspension behaviour and remain in contact with the road surface at all times. Although the vehicle responses are tracked and can be used e.g. for passenger comfort analysis, in this study, the primary role of the vehicle models is to generate realistic loading on the bridge. In addition, the vehicle dynamics tend to have a higher effect on the bridge responses when their natural frequencies match. In the case of very long-span bridges, the vehicle natural frequencies tend to be relatively higher, and most of the loading effects come therefore from the vehicle mass.

At each time step of the simulation, the coupled equations of motion for the bridge and all vehicles are solved. The interaction is enforced through contact constraints: each wheel experiences a contact force equal to the deflection of the tyre from its equilibrium, as well as the instantaneous velocity and acceleration. This depends on the difference between the wheel’s vertical position and the bridge deck surface deflection, along with the road roughness, at that specific point, ensuring that the vehicle and bridge motions influence each other. The coupled system is solved using a time integration scheme, yielding the vehicle forces $F_{\text{veh}}(t)$ on the bridge.

\subsubsection{Road roughness}

\begin{figure}[h]
\centering
\includegraphics[]{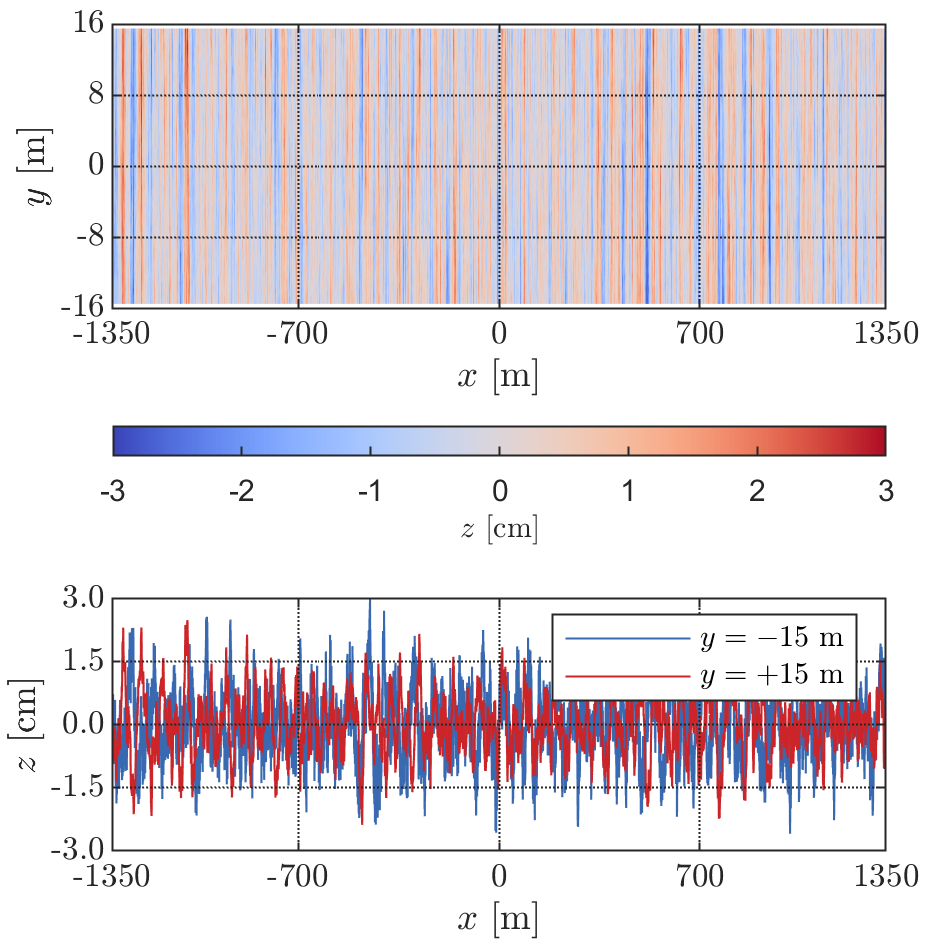}
\caption{Random road roughness according to ISO 8608, for a road of class B. Transverse profiles are correlated in time through spatial coherence functions.}
\label{fig:05}
\end{figure}

The road surface profile on the bridge is an important input for vehicle excitation and therefore bridge vibrations. A randomly generated road roughness profile is used, conforming to the ISO 8608 road roughness spectrum~\cite{iso8608}. The profile is defined by the vertical deviation $z(x,y)$ of the road surface from an ideal smooth reference along the longitudinal position $x$ and lateral position $y$, transverse across the lane. The PSD of the road roughness in the longitudinal direction depends on the road quality, assumed as ``B'' in this study to reflect a well-maintained structural condition. To account for the fact that the left and right wheel paths on a lane see slightly different road profiles but not entirely independent ones, transverse coherence is applied, analogous to the aerodynamic coherence~\cite{bogsjo2012}.

Figure~\ref{fig:05} illustrates the random roughness surface used throughout the length of the bridge, in longitudinal and transverse directions.

The profiles have maximum deviations of around 3 cm in this simulation. Notably, the two tracks exhibit similar long-wave trends but diverge at short intervals, reflecting the imposed partial coherence. This roughness serves as the excitation input to the vehicle suspension: as each wheel rolls, the relative displacement imposed on the vehicle is the bridge deflection plus the local road roughness at that wheel’s contact point.

\subsubsection{Traffic simulation}

\begin{figure}[h]
\centering
\includegraphics[]{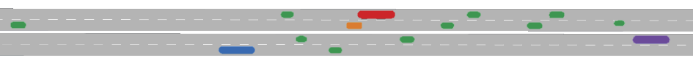}
\caption{Instantaneous configuration of a stochastic traffic simulation in VISSIM, showing different vehicle models (cars in red, vans in orange, buses in blue, unloaded trucks in violet and loaded trucks in red) in four different lanes (two per direction).}
\label{fig:06}
\end{figure}

The traffic loading is modelled stochastically to represent a realistic flow of vehicles crossing the bridge. Traffic composition is obtained from Danish databases~\cite{statbank2023}. In this study, we assume the highest recorded daily traffic volume for the Great Belt East Bridge of 56.000 vehicles/day, such that multiple vehicles are likely to be on the main span simultaneously. Vehicles enter each end of the bridge according to a probabilistic arrival process. Each arriving vehicle is randomly assigned a type based on the specified composition distribution. The vehicle’s parameters (mass, suspension properties) are then set according to its type, with random variability. The vehicle is also assigned an initial speed, and the deviation from the initial value throughout the simulation is controlled by a driving behaviour model, in this study the Wiedemann~74 method, implemented in the VISSIM software. 

Vehicles are assumed to travel in four lanes (two per direction) on the bridge. This leads to complex asymmetric loading scenarios. The time integration updates each vehicle’s position, until it reaches the end of the simulation domain. Figure~\ref{fig:06} provides an example snapshot of the traffic simulation in a portion of the bridge, indicating a mix of vehicles distributed along the span. This random traffic ensures that multiple vehicles may congregate on the span, increasing total load significantly, whereas at other times the span might be relatively lightly loaded.

\section{Dynamic analysis}  \label{sec:3}

\subsection{Simulation scenarios}

The numerical investigation considered four primary simulation scenarios:
\begin{enumerate}
    \item Scenario W: only turbulent wind excitation
    \item Scenario T: only traffic excitation
    \item Scenario W+T: linear supperpotision of the structural responses from scenarios W and T
    \item Scenario WT: turbulent wind and moving traffic simultaneously on the bridge
\end{enumerate}

These scenarios were designed to enable direct comparisons between isolated and coupled excitations. In particular, the responses from Scenarios W and T can be linearly superimposed and compared with the actual coupled response from Scenario WT, providing insight into the importance of interaction effects.

All simulations were carried out for a duration of 600~s, with traffic and/or wind applied for the whole simulation duration. The bridge was initially at rest, with vehicles entering the domain according to prescribed initial conditions. A short run-up period was included to minimise transient effects.

\subsection{Wind and traffic cases}

For consistency, each simulation employed one realisation of a turbulent wind field with a turbulence intensity of 10\%. Likewise, one realisation of the traffic stream was generated per case, consisting of randomised vehicle properties and flow patterns. The traffic conditions were defined by three allowable wind speed ranges~\cite{storebaelt2025}:

\begin{enumerate}
    \item Case 1: for low wind speeds (10-15 m/s), vehicles are limited to 110 km/h
    \item Case 2: in moderate wind speeds (15-20 m/s), the traffic limit is reduced to 90 km/h
    \item Case 3: for high wind speeds (20-25 m/s), a further limitation to 70 km/h is applied to traffic
\end{enumerate}

According to the bridge maintainer, in wind speeds higher than 25 m/s the bridge operation is stopped. These definitions ensure that the simulated scenarios reflect realistic operational restrictions under different wind climates.

\subsection{Results}

\begin{figure}[h]
\centering
\includegraphics[]{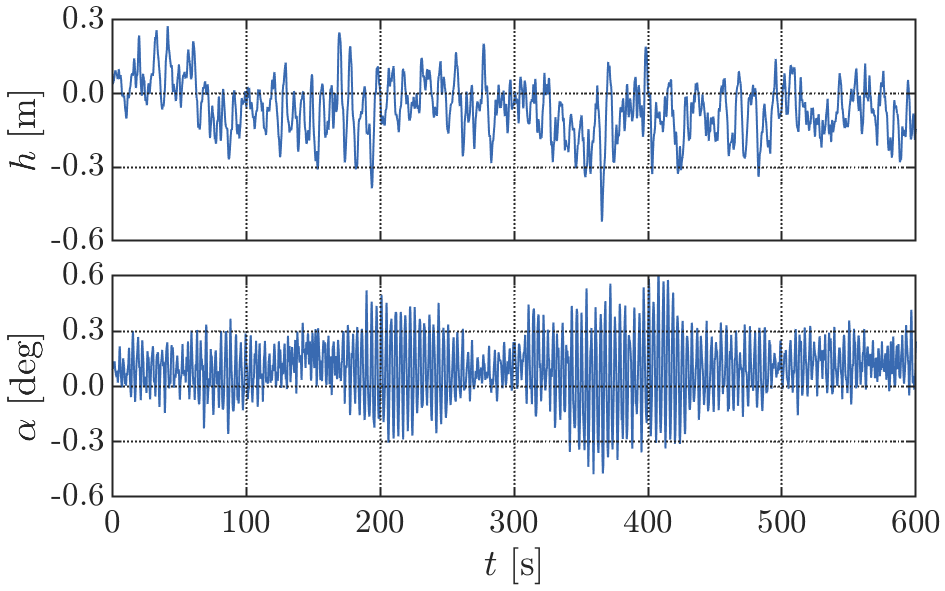}
\includegraphics[]{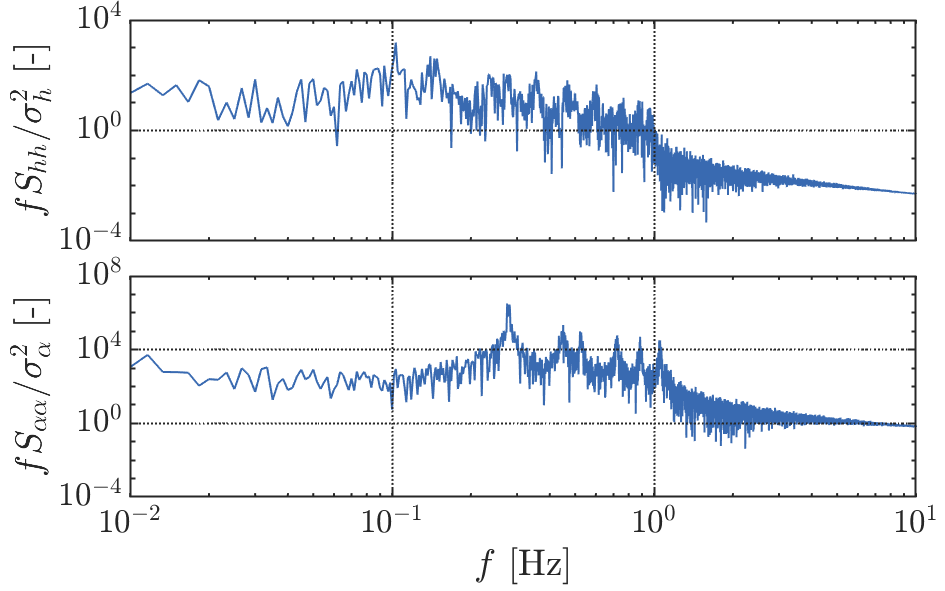}
\caption{Displacements and rotations (left) of the midspan of the bridge for an individual Case 3, Scenario WT analysis, along with the respective PSDs (right).}
\label{fig:07}
\end{figure}

The dynamic analysis of the model, for all scenarios and cases, returns time-series of displacements, velocities and accelerations of all structural and vehicle DOFs, while they were travelling on the bridge. In this work, the focus lies on the former, while the latter is not investigated. An example of the vertical displacements h and rotations $\alpha$ in the midspan of the bridge, for Scenario WT, along with the power spectral density (PSD) of the signals is shown in Figure~\ref{fig:07}.

In Scenario WT, wind and traffic act concurrently and interact non-linearly. Despite this coupling, dynamic amplification drives the response to align with the structure’s natural modes. Because the turbulent wind field, traffic stream, and road roughness are generated stochastically, results are evaluated via response envelopes over the entire structure, built from six independent realisations per scenario/case to account for stochastic variability.

\subsubsection{Different scenarios and linear superposition assessment}

\begin{figure}[h]
\centering
\includegraphics[]{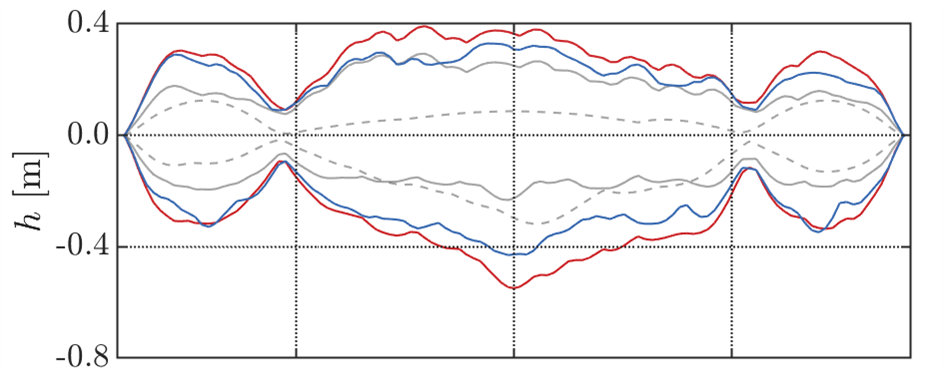}\\
\includegraphics[]{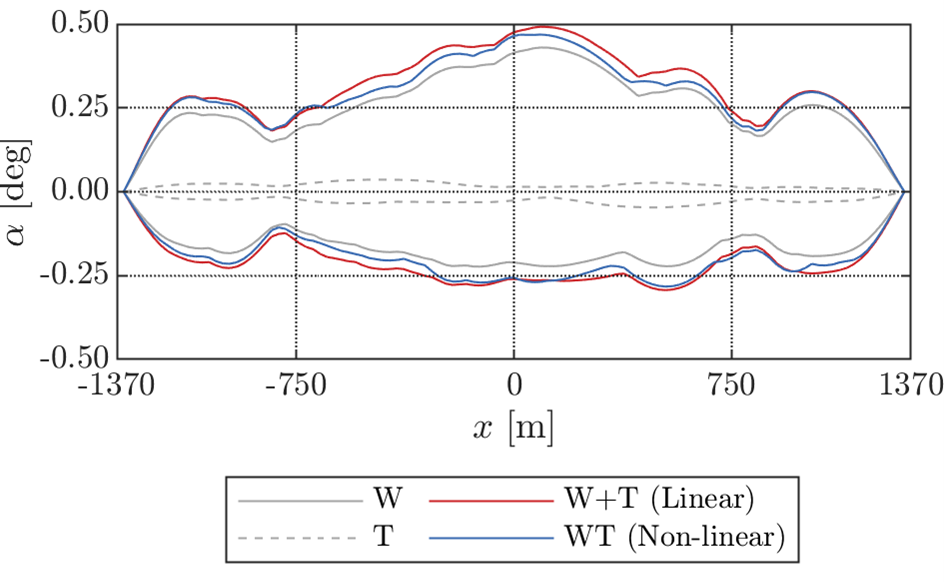}
\caption{Envelopes of displacements and rotations for Scenario W, Scenario T, a linear sum of W and T, and the non-linear Scenario WT.}
\label{fig:08}
\end{figure}

Initially, the bridge response under combined wind and traffic is compared to the superimposed response of separate simulations. Figure~\ref{fig:08} shows the deflection envelope along the span for four conditions: wind-only (Scenario W), traffic-only (Scenario T), a linear sum of scenarios W and T (W+T, linear), and the actual combined simulation, where wind and traffic act simultaneously on the bridge (Scenario WT). This analysis is shown for Case 3 (high wind speeds, low traffic speeds). 

The wind‐only envelope reflects buffeting and mean wind effects along the deck and is approximately symmetric about the undeformed state. The traffic‐only envelope exhibits larger downward than upward deflections, and has little rotation contribution in the deck, as expected for gravity-dominated loading. Linear superposition of the wind and traffic-only responses, commonly used in practice yields the red curve in Figure~\ref{fig:08} and predicts a mid-span deflection of around -0.5 m for this case.

When wind and traffic act concurrently, non-linearities arise from the coupled turbulent flow, moving vehicles, and structural dynamics. In scenario WT, the vertical response is slightly smaller in magnitude (at around -0.4 m at the centre of the span) the linear sum of scenarios W and T, mainly because the flow of vehicles breaks the correlation of the aerodynamic loads, which govern the structural response. The effect in rotations is similar, but less pronounced. 

The difference in response stems also from non-linear vehicle-bridge interaction: both vehicles and wind action can deflect the deck, as well as damp current structural vibrations based on the instantaneous traffic and aerodynamic configurations~\cite{camara2019}. Overall, Figure~\ref{fig:08} indicates that simple linear superposition may deviate from the coupled analysis results, as load distribution and timing are decisive. For design, treating wind and traffic separately (e.g., via envelopes) may miss critical simultaneous load cases or overpredict others, whereas direct time-domain simulation captures the coupled behaviour under realistic combined loading. In general, linear superposition is observed to be conservative, especially in the vertical responses, deeming code-based design appropriate but potentially not cost-effective.

\subsubsection{Wind and traffic interplay}

\begin{figure}[h]
\centering
\includegraphics[]{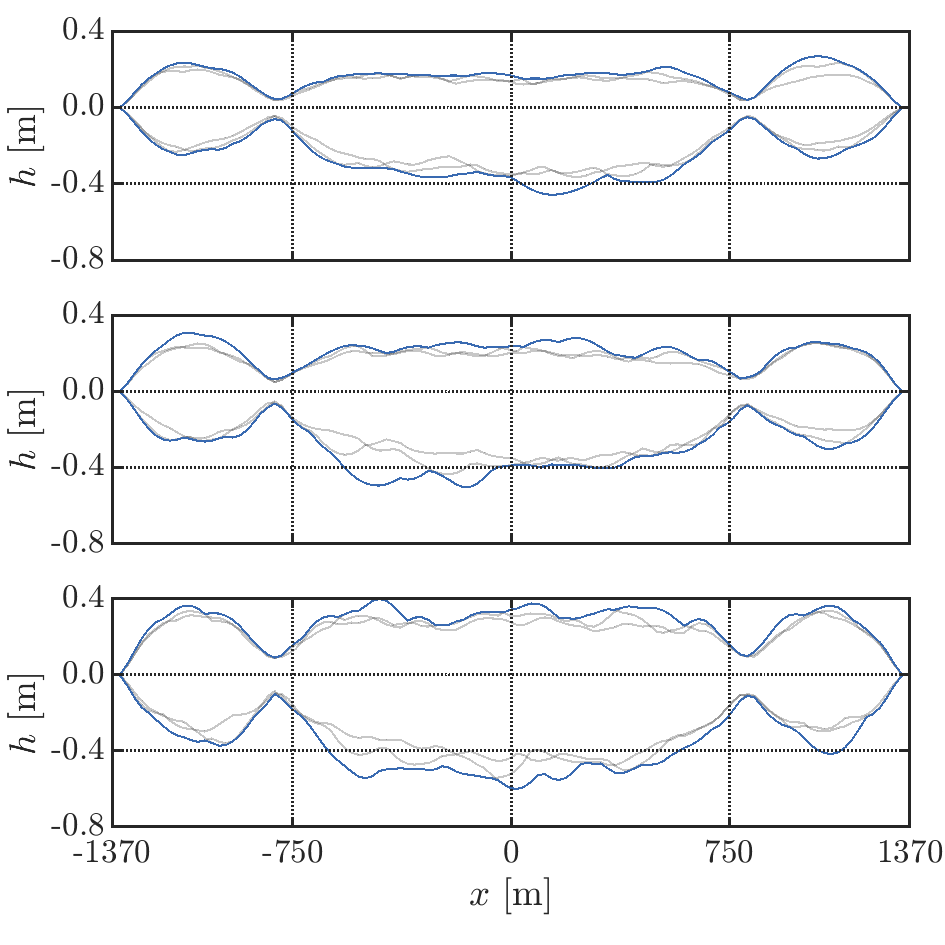}
\caption{Envelope of the three stochastic analyses for Scenario WT, for Case 1 (top), Case 2 (centre) and Case 3 (bottom). The blue line indicates the maximum between all envelopes.}
\label{fig:09}
\end{figure}

In the case of the Great Belt East Bridge, the traffic speed limitations are controlled by the current wind velocities on site. Figure~\ref{fig:09} shows response envelopes from three independent 10-min realisations of Cases 1, 2 and 3 (for Scenario WT), together with the envelope of vertical deflections across the realisations. In all cases, the maximum deflection occurs close to mid-span, as expected for a symmetric structural system. In Case 1 (fast traffic, low wind speed) the mid-span envelope attains approximately -0.45 m. In Case 3 (slow traffic, high wind speed) it increases to about -0.60~m, indicating that higher aerodynamic loading combined with slower traffic, and thus more simultaneous vehicles on the bridge, governs the response in the studied cases. The envelope shape shows a broad region of large deflection centred at mid-span (around 500 m). This region is broader and flatter for slower traffic, consistent with multiple heavy vehicles distributed along the span and producing near-maximum deflection over a wider zone. For faster traffic, the envelope is sharper, indicating more localised peak loading.

\section{Conclusions}  \label{sec:4}

In this work, the dynamic response of the Great Belt East Bridge under wind and traffic was analysed. A time-domain framework coupling a finite-element bridge model with a stochastic turbulent wind field and a microscopic traffic model, including random pavement roughness and vehicle-bridge interaction, has been developed.

Wind and traffic were found to interact non-linearly. Moving vehicles modify the effective instantaneous aerodynamic configuration, while wind-induced deck motions influence vehicle dynamics and force distribution. As a result, the combined response cannot be fully recovered by linear superposition of wind-only and traffic-only results. In most scenarios, linear superposition was conservative, largely because it implicitly assumes temporal coincidence of peak wind and traffic effects. However, at certain locations the non-linear combined response exceeded the linear sum, so simple addition is not reliably conservative.

Weather and traffic parameters were decisive. Slower or congested flows, especially under wind-induced speed restrictions, increase structural responses and load duration, elevating combined dynamic responses. Faster traffic generates shorter, more localised pulses. Therefore, the worst-case configuration is non-trivial and may arise from moderate wind coinciding with congestion rather than from isolated extremes.

The integrated stochastic framework effectively captured the coupled behaviour and supports detailed performance assessments. Although more computationally demanding than separate analyses, it is feasible for critical designs and in-depth studies. Design practice would benefit from explicit combined load cases, particularly for very long span bridges.

\bibliographystyle{unsrt}  
\bibliography{references}

\end{document}